\documentclass[review]{elsarticle}
\usepackage{hyperref}
\usepackage[margin=1in]{geometry}
\usepackage{amsmath}
\usepackage{amssymb}
\usepackage[version=4]{mhchem}
\usepackage{braket}
\usepackage{graphicx}
\usepackage{textcomp}
\usepackage{cleveref}
\usepackage{xcolor} % -- just for marking the text during the preparation
\usepackage{gensymb}
\usepackage{multirow}
\usepackage{epsfig}
\usepackage{graphicx} 
\usepackage{float}
\usepackage{makecell}
\usepackage{multirow}
\usepackage[normalem]{ulem} % to strikeout text
\usepackage{verbatim} % comment multiple lines
\usepackage{mathtools}  
\usepackage{xfrac}
\usepackage{ulem} %cross

\usepackage{colortbl} % -- to make tables with coloured cells 
\usepackage[switch, modulo]{lineno}

\usepackage[usestackEOL]{stackengine}

\usepackage{xr}
\journal{ArXiv}
\biboptions{sort&compress}

\begin{document}

%\setchemformula{bond-length=1em}

\begin{frontmatter}

	\title{Insights into Polymer Electrolyte Stability and Reaction Pathways: A first-principle calculations study\tnoteref{mytitlenote}}

	\author[affil:KZ]{Kazem Zhour\corref{mycorrespondingauthor}}
	\ead{kazem.zhour@uni-muenster.de}
	\address[affil:KZ]{Institut für Physikalische Chemie, Westf\"alische Wilhelms-Universit\"at M\"unster, 
	Correnssta{\ss}e 28, 48149 M\"unster, Germany}
	\cortext[mycorrespondingauthor]{Corresponding authors}
	
	\author[affil:KZ]{Andreas Heuer}
	\ead{andheuer@uni-muenster.de}

	\author[affil:DD]{Diddo Diddens}
	\address[affil:DD]{Helmholtz Institute M\"unster (IEK-12), Forschungszentrum J\"ulich GmbH, Corrensstra{\ss}e 46, 148159 M\"unster}
	\ead{d.diddens@fz-juelich.de}

\begin{abstract}

	This study investigates the electrochemical behavior and decomposition pathways of four monomers, namely PMC, PMC-OH, PeMC-OH, and DEO-EA, 
	which are potential candidates for polymer electrolytes in solid-state batteries. Density functional theory calculations were employed to 
	determine the oxidation and reduction potentials of these monomers near different ions ([Li]$^+$, TFSI$^-$, and [Li]$^+$[TFSI]$^-$) 
	and their corresponding reorganization energies. The results reveal notable sensitivity of the monomers to reduction in a [Li]$^+$-rich 
	regime and to oxidation in a TFSI$^-$-rich regime. Additionally, the decomposition pathways of the monomers were investigated, focusing 
	on the cleavage of CO bonds. The findings provide insights into the stability and reactivity of these monomers in various electrochemical environments.

\end{abstract}

\begin{keyword}
\texttt{density functional theory, solid state batteries, polymer electrolyte, solid electrolyte interphase}
\end{keyword}

\end{frontmatter}

\section{Introduction}

As lithium-ion battery technologies approach their energy density and safety thresholds \cite{song2023reflection}, there is an urgent need to 
seek replacements to accommodate the continuous increase in required energy. One highly promising candidate is the solid-state battery (SSB).
Such battery can push the energy density limit as it merges high-energy electrode material, such as Li metal anode and Lithium nickel manganese
cobalt oxides LiNi$_x$Mn$_y$Co$_{1-x-y}$O$_2$, know as NMC cathode, with solid electrolyte.

Among the candidates for solid electrolytes, polymer electrolytes (PEs) have attracted considerable 
attention due to their remarkable flexibility and compatibility with thin-film deposition techniques, as evidenced by referenced studies 
\cite{zhang2017single,lopez2019designing,mindemark2018beyond}. 
PEs offer a solution to the rigidity issue encountered with inorganic electrolytes, which often leads to physical contact problems 
and interface damage. While flexibility and thin-film compatibility are advantageous, the mechanical properties 
alone are insufficient for evaluating PEs. It advocates for considering a combination of properties, including lithium conductivity and 
electrochemical stability, to optimize performance.

A wide range of PEs with diverse designs and functional groups have been investigated 
with the aim of achieving desirable properties, particularly focusing on parameters such as ionic conductivity and transference number. 
These investigations delve into the detailed characterization of these properties to understand and optimize the performance of PEs.
For example, while the ionic conductivity of poly(ethylene oxide) (PEO) is slightly higher than that of poly(ethylene carbonate) (PEC), 
PEC demonstrates a much better transference number than PEO \cite{lopez2019designing}.
The importance of assessing the stability of PEs amidst electrochemical reactions, 
particularly at the interface, remains significant, despite the acknowledged challenge of ion conductivity in PEs, as noted in 
literature \cite{janek2023challenges}. This holistic evaluation approach indicates a nuanced understanding of PEs' 
functionality that extends beyond individual characteristics. In line with this, several recent experimental studies by Brunklaus and coworkers
have conducted extensive investigations on PEs beyond PEO \cite{mindemark2018beyond,chen2022green,imholt2019grafted,chen2023towards,chiou2023selection}.

To the best of our knowledge, there has not been an extensive, detailed theoretical investigation into these aspects for polymers
under different environments, 
unlike the substantial focus on solvent studies such as the detailed examination of ethyl carbonate (EC) and dimethyl carbonate (DMC) 
in lithium-ion batteries conducted by Borodin et al. \cite{borodin2015towards}, along with subsequent study for pyrazole-based additives \cite{von2020methyl}.
Recently, Zheng et al. \cite{zheng2021localized} employed ab initio molecular dynamics techniques to investigate potential reaction pathways 
following the reduction of a mixture containing 1,2-dimethoxyethane (DME) and tris(2,2,2-trifluoroethyl)orthoformate (TFEO) in various environments. 
This study show the efficiency of such technique in studying the electrochemical behavior and stability of electrolyte mixtures.
A recent publication by Marchiori et al. \cite{marchiori2020understanding} surveyed various polymer electrolytes, primarily focusing on studying 
the electrochemical window (EW) of the considered polymer without delving deeply into the consequent chemical reactions or checking the effect 
of different regions in batteries.
Additionally, Ebadi et al. \cite{ebadi2019assessing} investigated the reactivity of various polymers near an explicit Li anode, 
but again without considering the effect of different regions formed at the interface.

In general, the calculations conducted by Marchiori et al. reveal that the EWs significantly depend on the environment surrounding the electrolyte mixtures, 
where different slats were investigated. Specifically, they found that PEO near lithium bis(trifluoromethanesulfonyl)imide (LiTFSI) exhibits the widest 
EW with the lowest reduction potential, while polyethylenimine (PEI) displays the narrowest window.

Regarding the comparison between PEC and polyester poly($\epsilon$-caprolactone) (PCL) near Li TFSI$^-$, the study indicates that PEC demonstrates a wider 
EW with a higher oxidation voltage and a lower reduction potential than PCL. These findings highlight the importance of considering the specific environment 
in which electrolytes operate when assessing their electrochemical properties, providing valuable insights into the design and optimization of electrolyte
materials for various energy storage applications. Moreover, 
in batteries, various regimes can emerge where one type of ion dominates over the other. Generally, three distinct regimes can be identified: 
cation (e.g., Li$^+$) rich regime, anion (e.g., TFSI$^-$) rich regime, or an equivalent regime where both the cation and the anion are present 
in similar concentrations. Understanding and characterizing these different regimes are crucial for optimizing battery performance and stability

This observation underscores a potential gap in research that could benefit from further exploration, particularly concerning the electrochemical 
stability of PEs at the interface within different possible regimes. Closing this gap could provide valuable insights into optimizing the performance 
of PEs in various applications.

\section{Simulation details}
\label{sec:calcul}
\begin{figure}[!ht]
	\centering
	\includegraphics[width=0.3\linewidth]{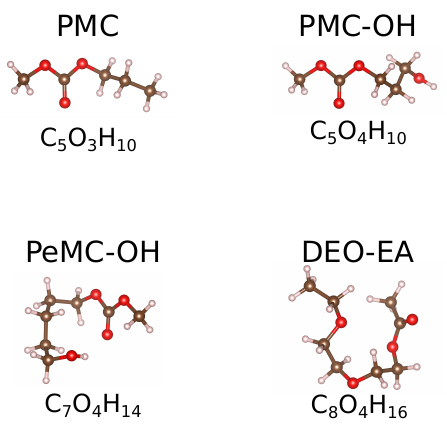}
	\caption{\label{fig:molecules}
	Molecular structures, abbreviations, and chemical symbols of the monomers employed in the research.
	The color code of the atoms is brown for carbon, red for oxygen, and white for hydrogen.}
\end{figure}

Four monomers: C$_4$O$_3$H$_{10}$ (PMC), C$_4$O$_4$H$_{10}$ (PMC-OH), 
C$_6$O$_4$H$_{13}$ (PeMC-OH), and C$_8$O$_4$H$_{16}$ (DEO-EA) as shown in Fig. \ref{fig:molecules} was examined in this work. 
The picked chemical name of such structures according to Syntelly \cite{syntelly} website are methyl propyl carbonate for PMC, 
hydroxypropyl methyl 
carbonate for PMC-OH, 4-hydroxypentyl methyl carbonate for PeMC-OH, and 2-(2-ethoxyethoxy)ethyl acetate for DEO-EA.
The selection of these monomers was intentional, covering a diverse range of CH$_2$ chain sizes, end group possibilities, and functional 
groups, including carbonate, ester, and ether functionalities.
Different conformers of these monomers were generated using CREST software \cite{pracht2020automated} from the extended 
tight binding (xtb) program package
based on semiempirical tight-binding methods combined with a meta-dynamics driven search algorithm. 
The most stable conformer for each monomer was chosen for the density functional theory (DFT) calculations.

The results for the cluster calculations of the four monomers near [Li]$^+$, TFSI$^-$, and [Li]$^+$[TFSI]$^-$
are obtained using DFT calculations with ORCA software \cite{neese2012orca,neese2022software}, where oxidation and
reduction potentials of all the considered monomers and their decomposition pathways have been calculated using
the long-range corrected hybrid density functional $\omega$B97X \cite{chai2008systematic} with Grimme's DFT-3 
\cite{grimme2010consistent} dispersion correction. The def2-TZVP \cite{weigend2005balanced} basis 
set was employed.
The conductor-like polarizable continuum model (CPMC) with THF solvent (relative dielectric constant $\epsilon_r = 7.25$) 
have been employed to mimic the environment of the electrolyte.

Molecular geometries were optimized, and frequency calculations were performed after each optimization.
In the presence of an imaginary frequency, a tight convergence criterion was applied for geometry optimization. 
Should the imaginary component persist, the optimized coordinates were modified along the eigenvector associated with 
that imaginary frequency, prompting a restart of the optimization process. Imaginary frequencies below 30 cm$^{-1}$ were disregarded, 
and structures corresponding to such cases were deemed stable.
Upon achieving the optimized geometry of a neutral molecule, subsequent calculations included both vertical and adiabatic oxidation 
and reduction analyses.

The reduction ($V_{\rm red}$) and oxidation ($V_{\rm ox}$) potentials for each monomer near [Li]$^+$, TFSI$^-$, and [Li]$^+$[TFSI]$^-$ 
were calculated according to the following equations respectively

\begin{gather}
	V_{\rm red} = - \frac{G(f^-)-G(f)}{F} - 1.4, \label{eq:E_red} 
\end{gather}
\begin{gather}
	V_{\rm ox} = \frac{G(f^+)-G(f)}{F} - 1.4, \label{eq:E_ox} 
\end{gather}

where $G(f^-)$, $G(f^+)$ and $G(f)$ are the free energies of reduced, oxidized and neutral monomer, 
$F$ is the Faraday constant. The subtraction of $1.4$V was applied to put the absolute values in context to the scale relative 
to the Li$\mid$Li$^+$ redox pair \cite{borodin2015towards,borodin2017modeling}.

The Gibbs free energy $G$ of each system in the gas phase at $T = 298.15$ K  is calculated as

\begin{gather}
	G = H - TS.
\end{gather}

The enthalpy $H$ is calculated from the inner energy $U$ as following

\begin{gather}
	%H = U + k_{B}T,
	H = U + pv,
\end{gather}

%where $k_{B}$ is the Boltzmann's constant, and

p is the pressure and v is the volume, and

\begin{gather}
	U = E_{\rm el} + E_{\rm ZPE} + E_{\rm vib} + E_{\rm rot} + E_{\rm trans},
\end{gather}

with $E_{\rm el}$ represents electronic total energy, $E_{\rm ZPE}$ is the zero-point energy, $E_{\rm vib}$ denotes the finite temperature correction 
to zero-point energy, $E_{\rm rot}$ stands for rotational thermal energy, and $E_{\rm trans}$ represents translational thermal energy.

The entropy $S$ is calculated as following

\begin{gather}
	S = S_{\rm el} + S_{\rm vib} + S_{\rm rot} + S_{\rm trans},
\end{gather}

where $S_{\rm el}$ represents electronic entropy, $S_{\rm vib}$ denotes the vibrational entropy,
 $S_{\rm rot}$ stands for rotational entropy, and $S_{\rm trans}$ represents translational entropy.

 The reorganization energy of reduced cluster $\lambda^{\rm red}$ and that oxidized $\lambda^{\rm ox}$  one were calculated according to Marcus theory
\cite{marcus1956theory} as follows,
\begin{gather}
	\lambda^{\rm red} = E^{\rm red}_{\rm ver} - E^{\rm red}_{\rm ad}, \label{eq:lambda_red} 
\end{gather}
where $E^{\rm red}_{\rm ver}$ is the energy needed to achieve a vertical reduction of the cluster by adding one electron without optimization, while
$E^{\rm red}_{\rm ad}$ is the adiabatic energy required to reduce the structure \ref{supp-figS:marcus-red}. Alternatively,
\begin{gather}
	\lambda^{\rm ox} = E^{\rm ox}_{\rm ver} - E^{\rm ox}_{\rm ad}, \label{eq:lambda_ox} 
\end{gather}
with $E^{\rm ox}_{\rm ver}$ is the vertical oxidation energy and $E^{\rm ox}_{\rm ad}$ is the adiabatic oxidation energy \ref{supp-figS:marcus-oxi}.

To study the decomposition of the monomers, particular attention was given to the carbon-oxygen bond within each monomer, 
evaluating the energy difference required for bond dissociation. In these calculations, both electronic energy and zero-point 
energy were considered to determine the overall energy in this context.

To model an electrified interface, an electron-rich regime of the batteries needs to be stimulated. 
Ab initio molecular dynamics (AIMD) simulations were performed to check the possible spontaneous decomposition of the polymers. 
This was done by adding an electron to the system every 1 ps for a total of 10 ps, considering the number of functional groups in 
the molecules and the CO bonds that require additional electrons to be broken. This methodology of including excess electrons in a 
cluster-based DFT calculation in the gas phase with implicit solvent and in periodic AIMD has been adopted previously in many studies  
\cite{wang2001theoretical,zheng2021localized,arora1998capacity,camacho2017elucidating}.
The r2SCAN-3c functional \cite{sun2015strongly,grimme2021r2scan} was employed. 
In our case, 
we added one electron every 1 ps because our simulation shows that all decomposition takes place within the first half of this interval. 
We tested one system by running it for 10 ps after adding two electrons, and no additional decomposition occurred. Additionally, 
adding 10 electrons to such a relatively small system may not be realistic. However, studying the initial few electrons can provide 
insight into the realistic part of the process, while the extra added electrons can be analyzed theoretically to understand what might 
happen if such conditions were ever met.
The Nosé–Hoover chain thermostat (NHC) was utilized with a coupling strength of 10 fs at a temperature of 300 K.

The Mulliken method \cite{mulliken1955electronic}, as implemented in Orca, was used to perform the charge analysis. 
Although this method assigns a partial 
charge to coordinated ions or molecules, reflecting the electron density redistribution due to coordination and the overlapping 
of basis functions for different atoms, it can still provide a good qualitative measure for extracting the charge of the formed fragments.

\section{Results and discussion}

\subsection{Electrochemical window} \label{para:EW}

\begin{figure}[!ht]
	\centering
	\includegraphics[width=0.75\linewidth]{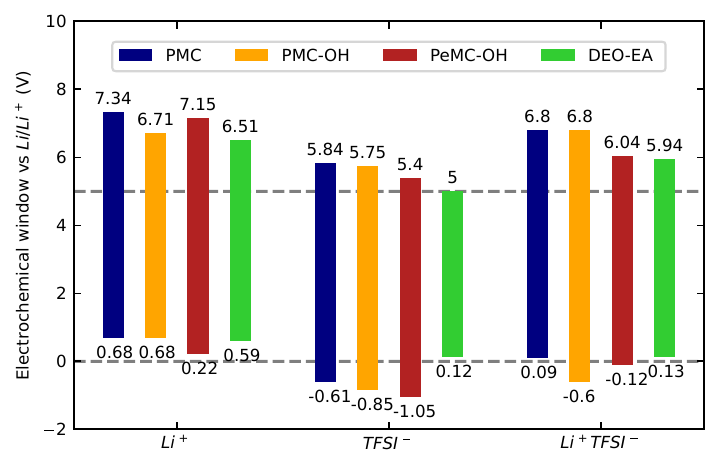}
	\caption{\label{fig:EW}
	EW of PMC, PMC-OH, PeMC-OH, and DEO-EA monomers in proximity to [Li]$^+$, TFSI$^-$, and [Li]$^+$[TFSI]$^-$.
	The dashed lines at 0 and 5 V correspond to the EW of  a typical lithium metal battery with an NMC cathode.
	The upper limits of the histogram represent the oxidation potential ($V_{\rm red}$) 
	while the lower ones denote the reduction potential ($V_{\rm ox}$).}
\end{figure}

The results of calculating the EW, derived from the oxidation and reduction potentials of each monomer 
in proximity to [Li]$^+$, TFSI$^-$, and [Li]$^+$[TFSI]$^-$ individually, are depicted in Figure \ref{fig:EW}, while the 
corresponding structures are presented in the Figure \ref{supp-figS:oxi-red}. From a general point of view, 
one can observe that the values of the oxidation and reduction potentials of all monomers near [Li]$^+$[TFSI]$^-$ are approximately 
the averages of the values near Li$^+$ and TFSI$^-$ separately, or at least fall in between them. An exception can be seen in 
the case of the oxidation of PMC-OH, where the oxidation potential of this monomer near [Li]$^+$[TFSI]$^-$ is even higher than 
that near Li$^+$. Upon checking the configurations, it is revealed that in all cases, the number of oxygen atoms coordinated 
to Li$^+$ remains the same before and after oxidation, except in the case of PMC-OH, where the coordinated oxygen reduces 
from three to two after oxidation.

The findings indicate a notable sensitivity of all the monomers to reduction in a [Li]$^+$ rich regime. Interestingly, 
increasing the number of the CH$_2$ groups, as observed in PeMC-OH, appears to diminish this sensitivity, while the addition of an 
OH end group does not exhibit a discernible effect. In the case of DEO-EA, the likelihood of reduction seems slightly less 
favorable compared to that of a single carbonate group. It is crucial to note that DEO-EA encompasses three functional groups 
(one ester and two ether), increasing its ability to coordinate with multiple [Li]$^+$ ions. Consequently, this multifunctionality raises 
the anticipation of a higher risk of reduction in such cases. To further investigate the effect of the number of functional groups, 
especially on the reduction potential, we consider DEO-EA without the tail, which than has only two functional groups, one ester and one ether, 
instead of three. In contrast, other monomers have only one group each. 
The calculated reduction potential near [Li]$^+$ rises slightly to become 0.65 V, while near TFSI$^-$ it reduces to -2.29 V, 
crossing the EW of the battery significantly, suggesting high stability of this monomer towards reduction near TFSI$^-$. Finally, 
near [Li]$^+$[TFSI]$^-$ it decreases slightly to -0.11 V. In general, the calculations suggest that the significant effect of the size of the 
monomer was observed in TFSI$^-$ rich regime, while for regimes containing [Li]$^+$, the effect was limited. 

Contrarily, in the TFSI$^-$ rich regime, all the monomers, except DEO-EA, exhibit a reluctance towards oxidation in addition 
to reduction. It is worth noting that monomers with shorter CH$_2$ chains demonstrate increased stability towards oxidation near 
TFSI$^-$—a contrast to their behavior near [Li]$^+$. This suggests that enhancing monomer stability in the cation-rich regime may 
come at the expense of stability towards oxidation in the anion-rich regime. The addition of an OH ending group still does not 
show a significant effect on the monomer's stability towards oxidation near TFSI$^-$ when comparing PMC to PMC-OH. Regarding DEO-EA, 
this monomer displays potential sensitivity towards both oxidation and surprisingly, reduction in such a regime.

In some of the double layers at the interface, both cations and anions can co-exist equivalently. 
The calculations in such a regime show that all the monomers exhibit stability toward oxidation, 
similar to the behavior observed in the cation-rich regime. 
However, challenges emerge in terms of reduction, where PMC and, as anticipated, 
DEO-EA show a propensity for reduction. Surprisingly, the addition of an OH ending group plays a crucial role in safeguarding 
the monomer in this scenario. Upon reviewing the structures, the [Li]$^+$ ions prefer to coordinate with 
both the monomer and TFSI$^-$ in the case of PMC-OH, whereas in the case of PMC, [Li]$^+$ coordinates only with TFSI$^-$—a distinction 
likely responsible for the noteworthy improvement in stability. 
Furthermore, another calculation was conducted for PMC, where [Li]$^+$ is coordinated to both the monomer and TFSI$^-$, 
revealing that the reduction potential decreases significantly to $-0.51$ eV, which is comparable to that of PMC-OH. Meanwhile, 
the oxidation potential shows no significant difference, with its value increasing by only $0.06$ eV. This further investigation suggests 
that the reduction potential is noticeably dependent on the arrangement of ions near the monomers. 

In the literature, it is widely acknowledged that TFSI$^-$ can exist in two distinct conformers: the trans conformer, characterized 
by a dihedral angle between the C-S bonds around $180^\circ$, and the cis conformer, with a dihedral angle approximately equal to $0^\circ$.
Table \ref{supp-tab:dihedral_TFSI} displays the evolution of the dihedral angle for neutral systems and after reduction and oxidation.
The optimized structures of all considered systems in this study consistently reveal that TFSI$^-$ adopts a trans conformation, 
maintaining a dihedral angle close to $180^\circ$, even after undergoing oxidation and reduction processes near all the monomers. 
An exception is observed in the case of DEO-EA, where the only case of rotation from trans to cis conformer is detected.
To delve deeper into the impact of the TFSI$^-$ 
conformer on reduction and oxidation potentials, a reevaluation of these processes near DEO-EA is conducted, specifically initiating the 
calculations from a trans conformer of TFSI$^-$. 
The results indicate that the reduction potential voltage becomes -0.99 V, which is comparable to the other monomers with the same conformer of TFSI$^-$. 
This suggests the higher reactivity of the cis conformer of TFSI$^-$ than the trans one. This aspect will be discussed in more detail in the section \ref{para:aimd}. 
The oxidation potential also increased to 5.85 V, making it equal to that of PMC. These new results suggest that the conformer of TFSI$^-$ was responsible 
for the distinction in the previous results of DEO-EA with respect to other monomers.

In order to explore the influence of monomer conformers on oxidation and reduction potentials, PMC was selected as a representative. 
The first nine most stable conformers of PMC near [Li]$^+$ were generated through tight-binding calculations using the CREST software, 
and their corresponding reduction and oxidation potentials were subsequently determined. The outcomes are illustrated in Figure \ref{supp-figS:confs}, 
while the optimized neutral structures are presented in Figure \ref{supp-figS:confs_configs}. Notably, the reduction potential is consistent across conformers, 
with a slight deviation observed in conformer number 8, distinguished by its coplanar C and O atoms. A similar trend is evident in the 
oxidation potential. Overall, it can be inferred that the influence of different conformers on oxidation and reduction potentials 
appears to be relatively minor.

\subsection{Kinetics} \label{para:Kinetics}

\begin{figure}[!ht]
	\centering
	\includegraphics[width=0.75\linewidth]{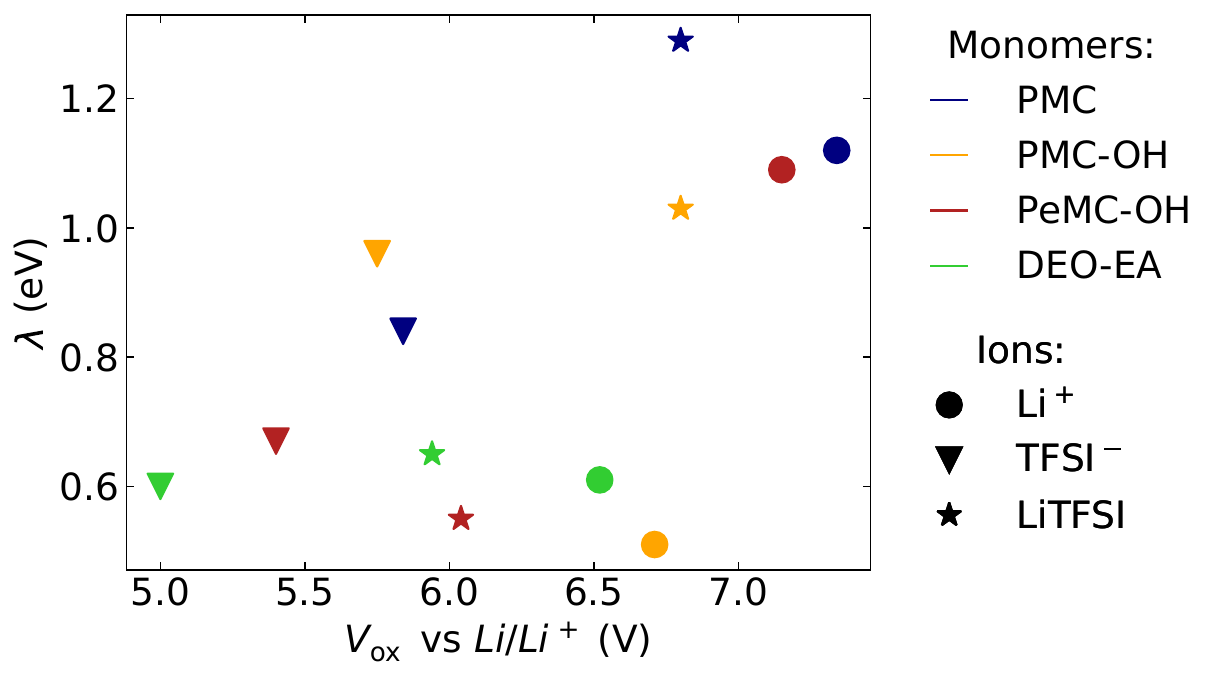}
	\caption{\label{fig:K-oxi}
	Reorganization energy ($\lambda$) for PMC, PMC-OH, PeMC-OH, and DEO-EA after oxidation near [Li]$^+$, TFSI$^-$, and [Li]$^+$[TFSI]$^-$ 
	as a function of oxidation stability ($G^{ox}_{ad}$).
	The lines represent fits near each ion separately.}
\end{figure}

\begin{figure}[!ht]
	\centering
	\includegraphics[width=0.75\linewidth]{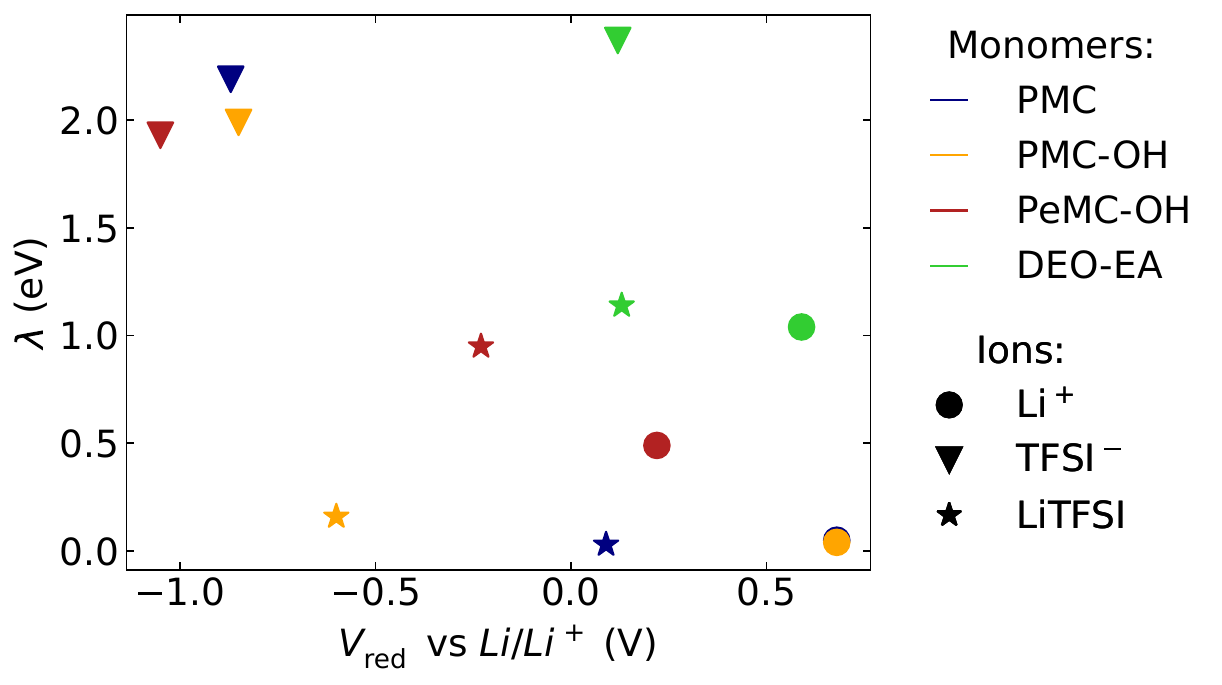}
	\caption{\label{fig:K-red}
	Reorganization energy ($\lambda$) for PMC, PMC-OH, PeMC-OH, and DEO-EA after reduction near [Li]$^+$, TFSI$^-$, and [Li]$^+$[TFSI]$^-$ 
	as a function of reduction stability ($G^{red}_{ad}$).}
\end{figure}

According to Marcus theory of electron transfer, the transfer rate is intricately linked to the reorganization energy ($\lambda$), arising from the disparity 
between adiabatic and vertical oxidation or reduction values separately (Eq. \ref{eq:lambda_red},\ref{eq:lambda_ox}). 
Figures \ref{fig:K-oxi} and \ref{fig:K-red} present the calculated $\lambda$ for each monomer near different ions after oxidation and reduction, respectively.
In general, the trend in the plots reveals that the value of $\lambda$ increases with the corresponding oxidation potential near each ion separately. 
This implies that monomers with higher oxidation potential exhibit a lower rate of electron transfer and, consequently, a decreased likelihood of decomposition. 
A linear trend has been experimentally confirmed for EC and DMC molecules \cite{egashira2001measurement}. 
Furthermore, a later simulation by Borodin et al. also validated a linear trend \cite{borodin2015towards} for the EC molecule in different
environment.
The divergence between $\lambda$ values depends on the ion type. For instance, near TFSI$^-$—where the oxidation potential is lower, as discussed in 
the previous paragraph—the difference in $\lambda$ values for different monomers is only $0.36$ eV. In contrast, near [Li]$^+$ and [Li]$^+$[TFSI]$^-$, 
which have higher oxidation potentials, the difference is $0.61$ and $0.74$ eV, respectively. Furthermore, the increasing slope of the lines depicting 
the trend near each ion suggests that, with an increase in oxidation potential, the electron transfer becomes even more challenging.
In terms of monomers, PMC consistently exhibits a lower transfer rate compared to other monomers near the same ion, while DEO-EA shows the highest. 
Surprisingly, The ranking of monomers with an OH ending group depends on the environment.

In the case of [Li]$^+$, PMC shows an exceptionally high reduction rate, with the value of $\lambda$ close to zero. Moreover, 
the addition of an OH ending group in the case of PMC-OH does not affect the value of ($\lambda$) 
(the data points of PMC and PMC-OH near [Li]$^+$ overlap in Figure \ref{fig:K-red}). This lack of effect on the reduction potential, 
as demonstrated earlier, leads to the consideration that such an ending group has no significant impact on the reduction rate of the monomer near [Li]$^+$.
On the other hand, increasing the number of CH$_2$ units in the monomer, as in the case of PeMC-OH, decreases both the reduction potential and the reaction rate.  
For other functional groups in DEO-EA, although the reduction potential remains comparable to that of PMC and PMC-OH, the reduction rate decreases 
significantly as $\lambda$ exceeds 1 eV in this case. This high value can also be attributed to the size of DEO-EA, which exceeds that of the others and involves 
multiple functional groups simultaneously.
A similar distribution of reduction rates can be observed near [Li]$^+$[TFSI]$^-$, where PMC and PMC-OH still show a very low values of $\lambda$ despite the 
difference between their reduction potentials. PeMC-OH and DEO-EA also exhibit high $\lambda$ near [Li]$^+$[TFSI]$^-$, similar to the case of Li$^+$.
In the case of reduction near TFSI$^-$, the trend is somewhat different. Adding the OH ending group relatively increases the reduction rate, 
and increasing the size of the monomer in PeMC-OH makes it slightly smaller. DEO-EA, however, maintains the highest $\lambda$ value compared to the other considered monomers.

Following the observation of a notably high reduction rate for PMC near [Li]$^+$ and [Li]$^+$[TFSI]$^-$, a more in-depth investigation was conducted by 
calculating the redox potential for gaining a second electron along with the corresponding $\lambda$. 
The results indicate a reduction in the redox potential of PMC near [Li]$^+$, reaching -0.05 V, thus maintaining its proximity to the edge of the battery's EW.
Simultaneously, $\lambda$ remains very low, with a value of 0.02 eV. These findings suggest not only the highly probable 
occurrence of the first reduction for PMC in the [Li]$^+$ rich regime with a high rate but also the high potential for a second reduction. In the case of [Li]$^+$[TFSI]$^-$, 
the redox potential of the second electron is -0.25 V, accompanied by $\lambda = 0.31$ eV, indicating a lower likelihood of such a transfer occurring.

\subsection{Decomposition} \label{para:decomp}

\begin{figure}[!ht]
	\centering
	\includegraphics[width=\linewidth]{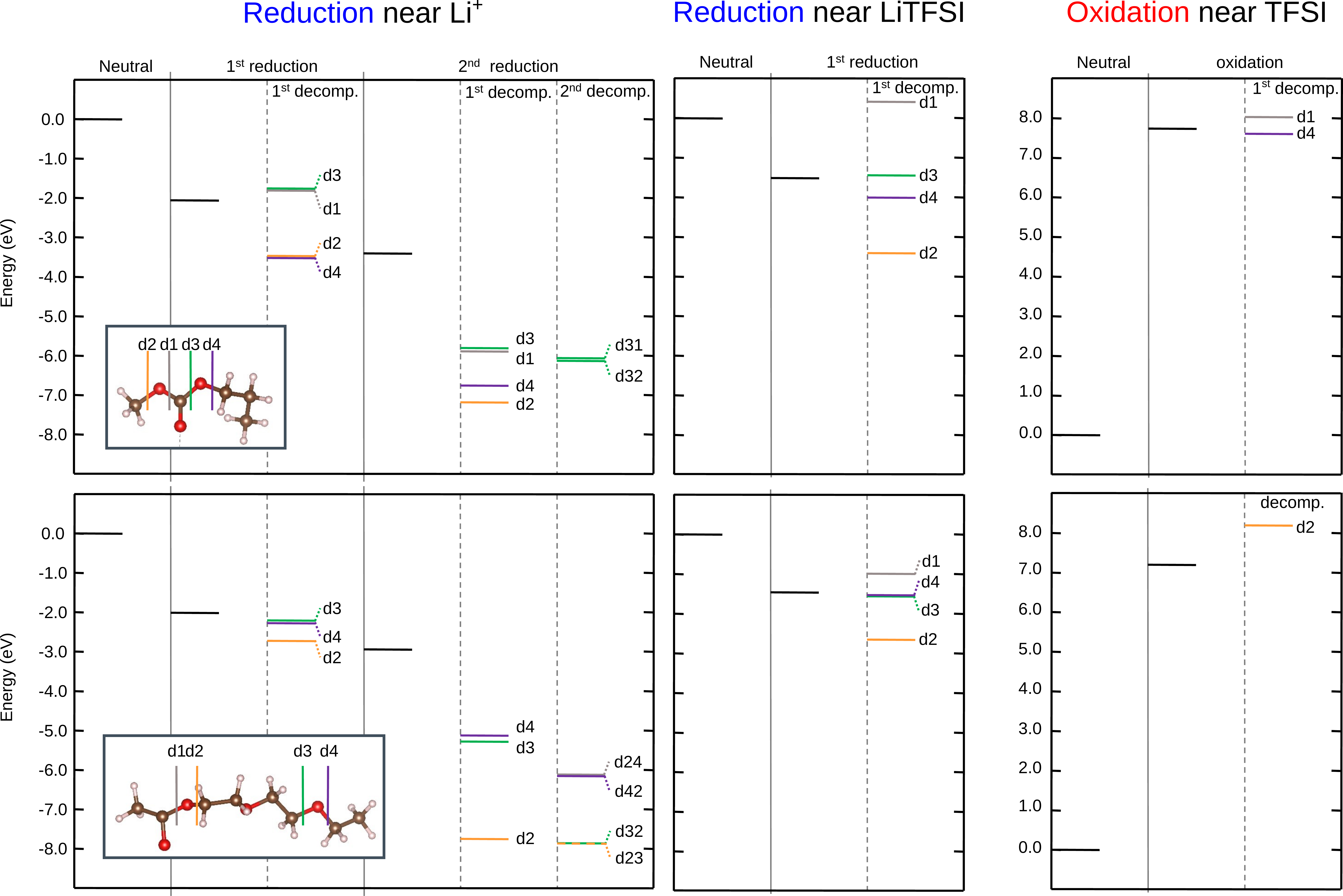}
	\caption{\label{fig:decomp}
	Energy diagram of the decomposition of PMC (top diagrams) and DEO-EA (bellow diagrams) after first and second reduction reduction near [Li]$^+$, 
	first reduction near [Li]$^+$[TFSI]$^-$, and oxidation near TFSI$^-$. A second version of these diagrams, showing the exact values of energies, 
	can be found in SI.
	The color code is similar to Fig. \ref{fig:molecules}.}
\end{figure}

After delving into the reduction and oxidation potentials, along with their respective reorganization energies in the preceding section, 
this section investigates the decomposition of monomers through the cleavage of CO bonds. In this analysis, particular attention is given to PMC, 
representing carbonate monomers, and DEO-EA. The investigation encompasses the decomposition pathways following the first and second reductions near [Li]$^+$, 
considering the high likelihood suggested by earlier calculations. 
Additionally, the decomposition only after the first reduction is explored near [Li]$^+$[TFSI]$^-$, 
as the electrochemical window (Figure \ref{fig:EW}) suggests that reduction of the monomer is highly probable to occur near [Li]$^+$[TFSI]$^-$, unlike oxidation. 
Moreover, decomposition following the first oxidation near TFSI$^-$ was also investigated.
Detailed structures resulting from reductions, oxidations, and all possible CO bond cleavages are provided in the 
SI. Figures \ref{supp-figS:F1_Li}-\ref{supp-figS:F1_TFSI} depict PMC structures, while Figures \ref{supp-figS:F3_Li}-\ref{supp-figS:F3_TFSI} represent DEO-EA. 
The corresponding energy diagrams for all investigated structures of PMC and DEO-EA are illustrated in Figure \ref{fig:decomp}. 

The CO bonds within the carbonate group of PMC can be categorized into three types: internal single CO bonds holding the carbonate group together, 
external CO bonds connecting the carbonate group to alkyl groups radicals (CH$_3$ and C$_3$H$_7$ in the case of PMC), and the double bond, 
which is beyond the scope of this study due to its expected higher strength. 
Upon the cleavage of CO bonds after the first reduction near [Li]$^+$ in the first two categories, a preference for breaking the external bonds 
(d2 and d4 in Figure \ref{fig:decomp}) is observed. There is a slight advantage in the formation of longer alkyl radicals C$_3$H$_7$ over the shorter CH$_3$, 
as the primary radicals are expected to be less stable than the secondary ones. 
Conversely, the cleavage of internal bonds (d1 and d3) is less likely to occur, as their energy is slightly higher than that of the reduced monomer. 
In contrast to alkyl radicals, longer alkoxy radicals (C$_3$H$_7$O) are slightly less likely to be formed than the shorter (CH$_3$O) ones in this case. 
These results suggest the dominance of alkyl radical formation, 
capable of combining to later form alkane gas, over the formation of alkoxy radicals. 
Notably, all types of these radicals have been experimentally detected \cite{wu2021recent,xu2006syntheses,hobold2020operando,eshetu2018ultrahigh}. 
After the second reduction near [Li]$^+$, both types of CO bond cleavage become energetically favorable. 
In a second bond cleavage, only d31 and d32 could survive (refer to \ref{supp-figS:F1_Li}) and exhibit a slight preference over the initial decomposition. 
The d13 and d31 pathways lead to the formation of CO and CO$_2$, occurring only after the second reduction of the PMC monomer. 
It is essential to note that we are not calculating the energy barrier for such decompositions here; instead, we are assessing the preference between 
structures before and after bond breaking.

Similar to the [Li]$^+$ rich regime, the cleavage of external bonds exhibits preferences over the reduced PMC, while internal bonds are less favorable near [Li]$^+$[TFSI]$^-$. 
After d3 and d4, the energies are slightly higher and lower than reduced PMC, 
respectively, than the reduced PMC. Following d2, the energy of the structure becomes significantly lower, indicating a high likelihood of CH$_3$ radical formation after 
reduction near [Li]$^+$[TFSI]$^-$. In contrast, after d1, the energy increases by 1.93 eV, suggesting a low possibility of CH$_3$O formation in such a regime.
The broad distribution of energy for different 
bond breaking events, unlike the case of [Li]$^+$ rich regime, can be attributed to the re-coordination of Li$^+$ ions with TFSI$^-$ and the formed radicals 
(refer to Figure \ref{supp-figS:F1_LiTFSI} and Table \ref{supp-tab:F1_LiTFSI}). For instance, CH$_3$O is pushed away from the first coordination shell of Li$^+$
after bond decomposition d1, while the alkoxy radical C$_3$H$_7$O remains coordinated to Li$^+$ after d3.
Another calculation was performed after the d1 decomposition, where the CH$_3$O was pushed to coordinate with Li$^+$. Optimizing the system 
led to the formation of a cluster with an energy of $-$1.59 eV relative to the neutral undecomposed structure, which is more stable than the cluster 
with uncoordinated CH$_3$O.

After oxidation near TFSI$^-$, only d1 and d4 decompositions were identified by DFT calculations. In both cases, a spontaneous proton transfer occurs between monomer 
fragments after system relaxation. d1 is slightly less favorable than oxidized PMC, and resulting in the formation of CH$_2$O due to the hydrogen transfer. 
Conversely, d4 is slightly more favorable and leads to the formation of propene (C$_3$H$_6$), also facilitated by hydrogen transfer.

As mentioned earlier, DEO-EA features two types of functional groups: one ester and two ether groups. The cleavage of single CO bonds from both the ester 
and ether group at the edges was considered. The energies of the corresponding structures before and after cleavage are displayed in Figure \ref{fig:decomp}. 
After the first reduction, the breaking of the internal CO bond in the ester group (d1) did not survive, as this bond tends to recombine after the optimization 
of the broken structure. On the other hand, decomposition d2 from the ester group, leading to the formation of C$_2$H$_3$O$_2$, exhibits a higher preference 
over both bond cleavages of the ether group (d3 and d4). Similar to PMC, the second reduction of DEO-EA significantly increases the preference for all bond cleavages, 
with d1 consistently holding the lowest energy compared to other bond cleavages. Notably, after the relaxation of the second reduction following d2, 
there is spontaneous formation of C$_2$H$_4$, indicating that this reaction has no energy barrier to occur (Fig. \ref{supp-figS:F3_Li}). 
The survived second decompositions are also presented in the SI, where all of them, except d24, exhibit a preference over their corresponding original structures. 
These decompositions result in various fragments such as C$_2$H$_5$, C$_2$H$_5$O, and C$_2$H$_3$O$_2$.

Contrary to the Li$^+$ rich regime, d1 survives after reduction near [Li]$^+$[TFSI]$^-$ and leads to the formation of C$_2$H$_3$O. However, the system has a 
higher energy level than the reduced DEO-EA, making it less probable to occur than other decompositions that result in structures with lower energies than the 
undecomposed DEO-EA. Similarly, d2 leads to lowest energy, and d3 and d4 have very close energies.

Finally, after the oxidation of DEO-EA in a TFSI$^-$ rich regime, only d2 survives, but it leads to a structure with an energy level approximately 1 eV higher 
than the undecomposed oxidized monomer. This, together with the high oxidation potential, suggests the rarity of witnessing decomposition of the monomer in such conditions.

\subsection{Spontaneous reaction} \label{para:aimd}

After studying the stimulated CO bonds cleavages in the previous section, a spontaneous method based on AIMD will be discussed here. In this method, 
one electron will be added after every one picosecond of MD run. 
Although this electron transfer rate significantly exceeds the experimental current density, 
a high rate can be expected at the beginning of the formation of the solid electrolyte interface. Additionally, 
due to computational limitations, considering a lower rate would result in a much higher computational cost.
In total, ten electrons are added to each system within 10 ps 
of simulation time, following a protocol similar to that used by Zheng et al. \cite{zheng2021localized}. 
Using this protocol, an oligomer composed of three PMC monomers and the DEO-EA monomer are investigated in this section.

The oxidation and reduction potentials of the oligomer near Li$^+$ are found to be 7.38 V and 0.75 V, respectively, while those near 
[Li]$^+$[TFSI]$^-$ are 6.66 V and -0.42 V, respectively. The corresponding reorganization energies are 0.08 eV, 0.09 eV, 0.96 eV, and 0.44 eV, 
following the same previous order. These values are somehow different from those of the PMC monomer 
(refer to Figures \ref{fig:EW}-\ref{fig:K-red}), 
which suggests that the electrochemical window and the corresponding kinetics depend on the oligomer size, 
its termination (CH$_3$ in the present case) or its conformer.

\begin{figure}[!ht]
	\centering
	\includegraphics[width=\linewidth]{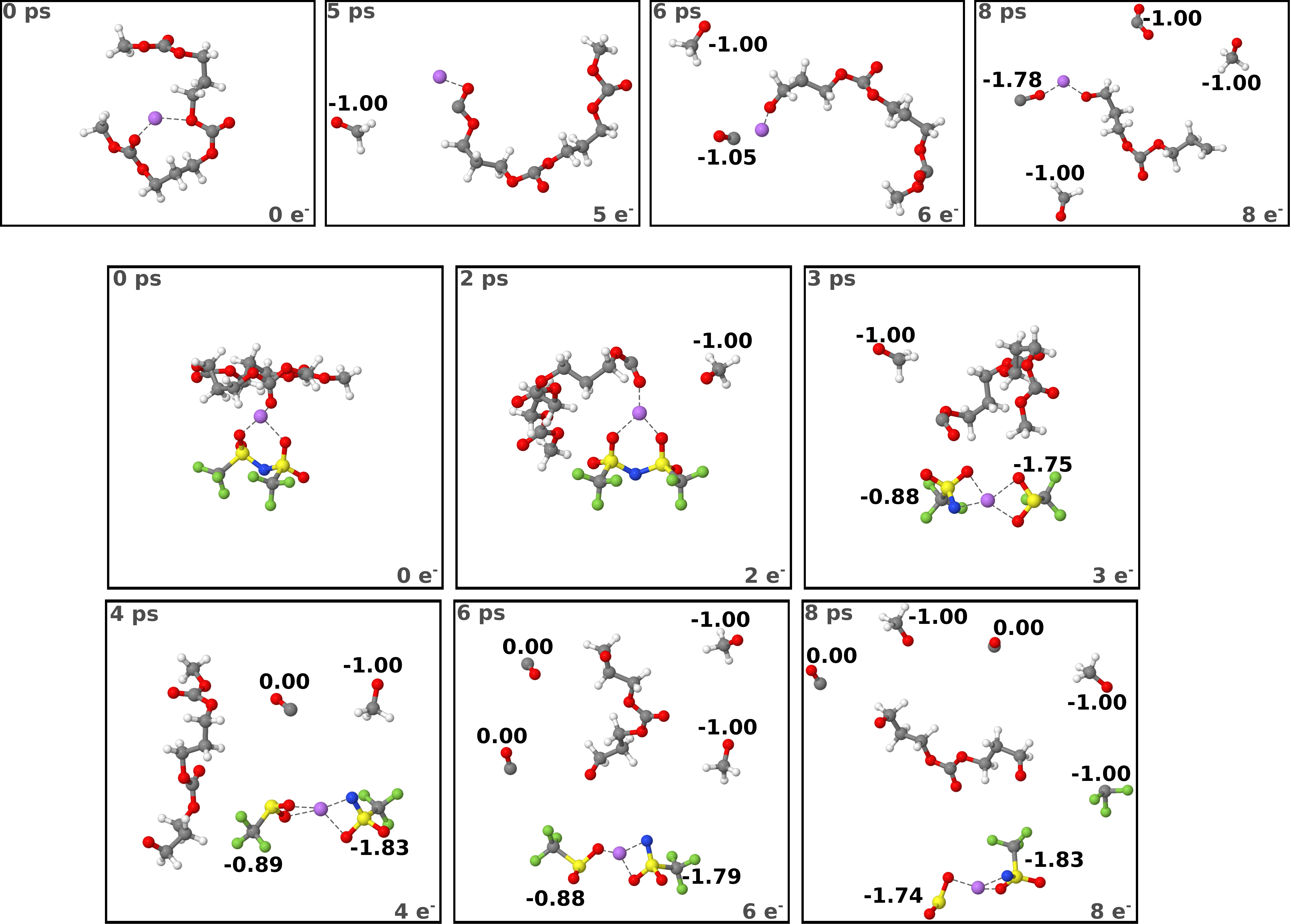}
	\caption{\label{fig:AIMD-PMC}
	PMC reaction mechanism diagrams near [Li]$^+$ (top diagrams) and [Li]$^+$[TFSI]$^-$ (below diagrams) while running AIMD simulations 
	for 1 ps after adding each electron. 
	Charges of the produced fragments are given in |e|.
	The dashed line shows the coordination of the atoms in the system with Li.
	The color code is: grey for carbon, red for oxygen, white for hydrogen, purple for lithium, blue for nitrogen, yellow for sulfur, 
	and green for fluorine.
	}
\end{figure}

As shown in the upper panel of Figure \ref{fig:AIMD-PMC}, no decomposition of the PMC oligomer  occurred on the timescale probed by the AIMD simulations
 before adding the fifth electron. The Mulliken charges show that the first four electrons added to PMC were primarily 
localized on the three carbonate groups and the Li ion, with approximately one electron on each. This localization may explain why no rapid 
decomposition of the polymer was observed unless more than one electron is localized per carbonate group. This situation occurred after 
adding the fifth electron (Fig. \ref{fig:AIMD-PMC}), which resulted in the detachment of CH$_3$O$^-$ from the carbonate group coordinated to Li. 
The calculated energy difference in the previous section suggests that the detachment of the CH$_3$ radical has a lower energy. 
This disparity mainly arises from the fact that the molecules in Figure \ref{fig:decomp} and \ref{fig:AIMD-PMC} have different potential energy 
landscapes (Born-Oppenheimer surfaces) due to the varying number of electrons involved in bond breaking in each case. 
Additionally, this result may also indicate that the activation energy for the detachment of CH$_3$O is lower than that of CH$_3$, 
which directly impacts the reaction rate of the system according to the Arrhenius model.
Adding the sixth electron led to the breaking of another CO bond from the same carbonate group, resulting in the formation of a CO$^-$ fragment.
Additional investigation on the state of the system after adding the sixth electron has been performed by comparing the energy of 
the same configuration at 6 ps but with  5e$^-$ instead of  6e$^-$ to the structure at 5 ps with  5e$^-$. This side calculation shows that the configuration at 6 ps, 
where CO is detached from the rest of the oligomer but with  5e$^-$, is energetically preferred by 0.49 eV over the configuration at 5 ps where the 
CO is not yet detached. 
This result suggests that if the simulation runs for a sufficiently long period after adding the fifth electron, the decomposition of CO will occur without 
the need for adding the sixth electron. However, achieving such a condition may require an extended period of time, which would be computationally very expensive.
Moreover, the reduction potential for adding the sixth electron was found to be -0.73 V vs Li/Li$^+$, suggesting that the reaction mechanism 
without the sixth electron might be more plausible, even if it takes more time.
Upon continuous reduction, additional CH$_3$O$^-$ fragments (or longer alcoholates) are formed.
These observations indicate that the decomposition of PMC near Li$^+$ is highly dependent on the localization of electrons within the molecule. 
The results suggest that a higher density of localized electrons on specific carbonate groups is necessary to induce rapid decomposition. 

For PMC near [Li]$^+$[TFSI]$^-$, after adding the second electron, the CH$_3$O$^-$ fragment detached (Lower part of Fig. \ref{fig:AIMD-PMC}). 
This detachment occurred due to the breaking the carbonate group coordinated to Li, 
suggesting that the decomposition of such a polymer near [Li]$^+$[TFSI]$^-$ is more vulnerable to reduction than when it is in the vicinity of the 
Li$^+$ ion alone. 
This result seems to conflict with the findings shown in the previous section in Figure \ref{fig:decomp}, 
where the decomposition of CH$_3$O$^-$ is shown to be more favorable near Li$^+$ than near [Li]$^+$[TFSI]$^-$. 
However, several differences between the two systems account for this discrepancy. 
Firstly, in the AIMD case, two electrons were added to the system, while the static calculations considered only the first reduction near  TFSI$^-$. 
Moreover, in the AIMD simulations, the Li ion is sandwiched between the oligomer and the  TFSI$^-$, whereas in the static calculations, 
it was coordinated to TFSI$^-$ without the PMC monomer (Fig. \ref{supp-figS:F1_LiTFSI}).
This sandwich coordination of the Li ion seems to play a major role in 
localizing both added electrons on the coordinated carbonate group, leading to its decomposition. In contrast, when the oligomer 
is in the vicinity of Li$^+$ alone, the electrons prefer to be distributed among all the carbonate groups, as discussed previously.
Finally, TFSI bears another negative charge, rationalizing that fewer extra electrons are needed to induce decomposition.
Adding the third electron led to the break of the SN bond in TFSI$^-$, forming SO$_2$CF$_3^-$ and NSO$_2$CF$_3^{2-}$ 
fragments, as indicated by Mulliken charge analysis. The Li$^+$ ion completely left the molecule and coordinated only with the fragments 
of TFSI$^-$. When the fourth electron was added, a neutral CO molecule detached from the molecule by breaking the CO bond of the 
same previously decomposed carbonate group. The additional electron localized mainly on the remaining part of the molecule, which 
became negatively charged.
Adding the sixth electron generated another CH$_3$O$^-$ and a CO molecule by decomposing the carbonate monomer on the opposite edge of 
the remaining PMC molecule. The eighth electron led to the decomposition of SO$_2$CF$_3^-$ into SO$_2^-$, 
which might rather easily receive another electron to form sulfoxylate,  and CF$_3^-$ fragments. 
The last two electrons surprisingly localized on one of the CO molecules,
therefore, these highly reduced clusters were deemed unphysical and were not investigated further. 
This case requires additional investigation to determine the reason behind such results.	
These observations indicate that the decomposition of PMC near [Li]$^+$[TFSI]$^-$, similar to near [Li$]^+$, generates CH$_3$O$^-$ fragments, 
but also neutral CO molecules, which are more stable than CO$^-$. Additionally, it is notable that TFSI$^-$ is
less vulnerable than the PMC polymer reductive decomposition.

 \begin{figure}[!ht]
	\centering
	\includegraphics[width=\linewidth]{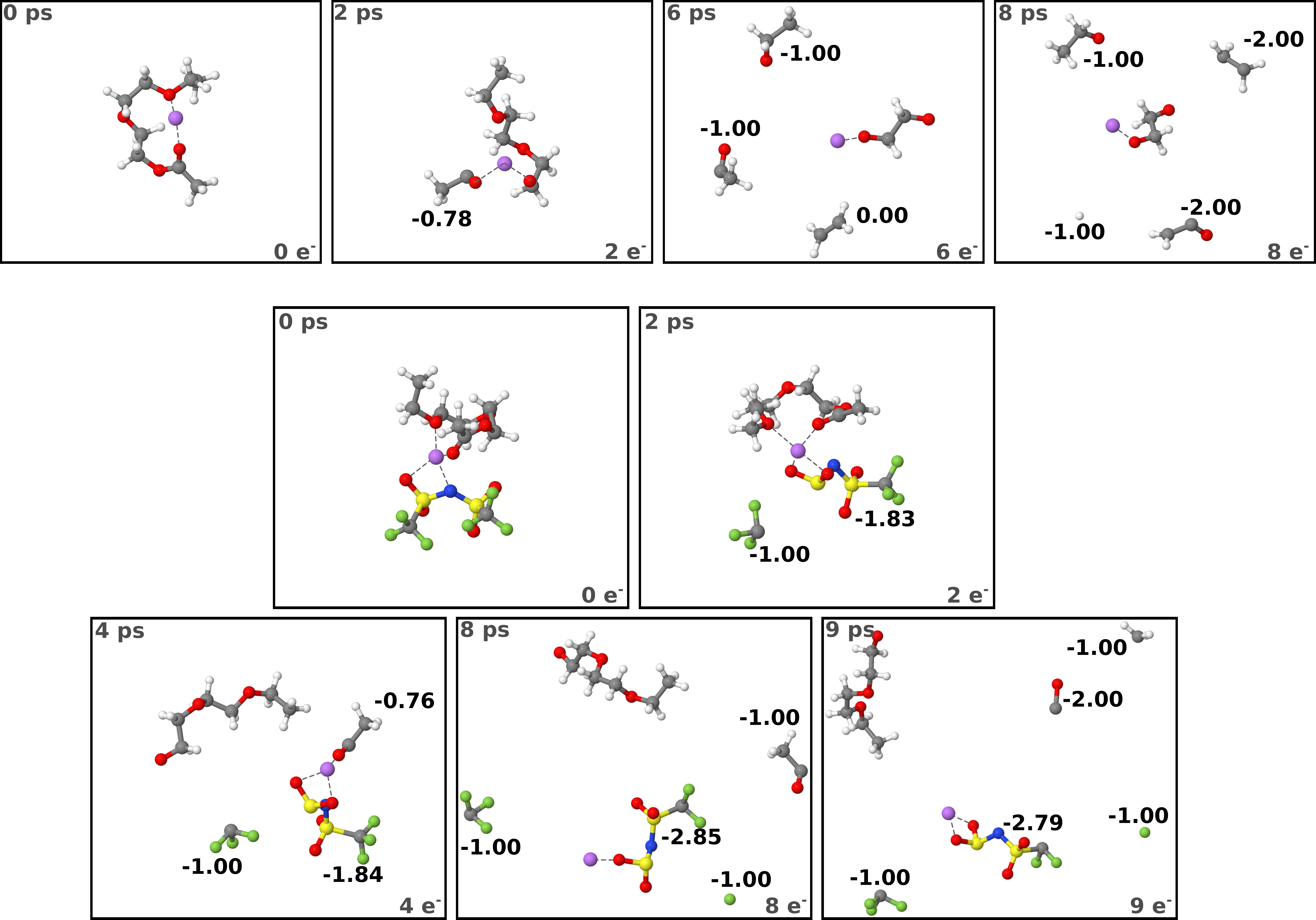}
	\caption{\label{fig:AIMD-DEO-EA}
	DEO-AE reaction mechanism diagrams. All details are similar to the previous figure.}
\end{figure}

For the DEO-EA monomer, the addition of two electrons in the [Li$]^+$-rich regime was sufficient to break the ester group (Fig. \ref{fig:AIMD-DEO-EA}), 
leading to the release of the C$_2$H$_3$O$^-$ fragment. This cleavage was unstable in the manual cleavage discussed earlier, 
which involved only one electron. However, upon optimizing the system after adding two electrons at once, it led to a stable 
structure with an energy lower than the neutral undecomposed system by 4.50 eV, which indicates a spontaneous reaction, as 
4.5/2 - 1.4 = 0.85 V vs Li/Li$^+$ falls within the electrochemical window. However, the formed C$_2$H$_3$O$^-$ fragment would be rather 
short-lived and could, for example, grab a proton from somewhere. 
Additionally, this decomposition suggests that the ester group is less stable than the ether group, especially when they coexist in 
the same polymer. This result will be further confirmed later when considering different starting configurations.
Upon adding six electrons, both ether groups broke, 
resulting in the formation of another C$_2$H$_3$O$^-$ fragment and a C$_2$H$_4$ molecule. 
The calculations show that the same configuration with  5e$^-$ is 4.50 eV more favorable than the configuration given at 5 ps. 
Therefore, similar to the case of PMC, such decomposition may take place without the need for adding the sixth electron if the simulation 
runs for an extended time. However, the reduction potential calculation found to be 3.61 V vs Li/Li$^+$, demonstrates the capability of 
the system to capture more electrons.
Upon the addition of the eighth electron, a hydride (H$^-$) was launched from the first C$_2$H$_3$O fragment at this moment, which would 
react rapidly. 

In the [Li]$^+$[TFSI]$^-$ environment, initially, the CF$_3^-$ group detached from TFSI$^-$ after the addition of two electrons
(Fig. \ref{fig:AIMD-DEO-EA}). 
With two more electrons, the ether group broke, forming the C$_2$H$_3$O$^-$ ion, similar to the [Li$]^+$-rich regime. 
It was not until the addition of the eighth electron that further decomposition occurred, where an F$^-$ ion was detached from 
the remaining fragment of TFSI$^-$. This F$^-$ can contribute directly to the formation of LiF components of the solid electrolyte 
interphase. 
Further investigation shows that the formation of the F$^-$ ion can occur with only 7 electrons if the AIMD simulation runs for a sufficient time. 
Single-point calculations of the configuration at 8 ps with 7e$^-$ are more favorable than those with 8e$^-$ by 0.15 eV. The same test was conducted with 6e$^-$, 
but the calculations indicate that this number of electrons may not be sufficient to observe the generation of the F$^-$ ion, as the energy difference favored the '
8e$^-$ configuration by 0.58 eV.
In general, the results suggest that TFSI$^-$ is more vulnerable to decomposition than DEO-EA, a case that will be discussed 
in more detail in the next paragraph.

One important aspect to check is the effect of the initial configuration on the decomposition path of the monomer. For this purpose, we considered the case of DEO-EA in 
the proximity of [Li]$^+$[TFSI]$^-$. In addition to the previously studied configuration, we performed a molecular dynamics simulation run for 2 ns based on 
the tight-binding model GFN2-xTB \cite{bannwarth2019gfn2}. 
The two configurations produced after each 1 ns were considered as initial configurations for repeating the AIMD studies, 
similar to the previous cases (Fig. \ref{supp-figS:AIMD_F3_LiTFSI}). 
Its important to note that after 1 ns, TFSI$^-$ maintains its cis conformer, although the coordination of the Li$^+$ changes. However, after 2 ns, 
the conformer of TFSI$^-$ switches to trans.
In general, the key observation is that the decomposition path clearly depends strongly on the initial configuration. Also, for these limited examples, 
one can notice that the cis conformer of TFSI$^-$ tends to be more reactive than the monomer, while in the case of the trans conformer, DEO-EA monomer decomposes 
initially.
This is most probably related to the fact that the trans conformer of TFSI$^-$ is more stable than the cis conformer 
(\cite{herstedt2005spectroscopic,lassegues2009spectroscopic}).
It is important to mention that the first decomposition of the DEO-EA always occurs by the breakage of the CO bond of the ester group, 
confirming the previous suggestion that the ester group is less stable than the ether group.

\section{Conclusions}

In summary, this study comprehensively examined the electrochemical properties and decomposition behavior of four monomers proposed for polymer 
electrolytes in solid-state batteries. The investigation revealed distinct sensitivities of the monomers to reduction and oxidation depending 
on the ion-rich environment. PMC exhibited high reactivity towards reduction near [Li]$^+$, while DEO-EA showed sensitivity towards both reduction 
and oxidation near [Li]$^+$ and [Li]$^+$[TFSI]$^-$, respectively. The decomposition pathways highlighted the preference for cleavage of external 
CO bonds in carbonate monomers, leading to the formation of CO and CO$_2$ molecules together with other fragments and components.
The required number of electrons to observe decomposition in the polymer depends on the number of functional groups present. 
This is evident in the case of PMC with three carbonate groups, where five electrons were necessary to induce decomposition using AIMD.
Spontaneous reactions consistently demonstrate that the functional group coordinated to the Li ion exhibits a preference in decomposition over 
other similar or distinct functional groups that lack a nearby Li ion.
AIMD calculations also suggest that the ester group is less stable than the ether group and is the first to decompose when they coexist in the same polymer.
Moreover, the decomposition order and path depend significantly on the initial configuration of the system. For example, the cis conformer of 
[TFSI]$^-$ tends to be more reactive than the monomer, while the trans conformer results in the initial decomposition of the DEO-EA monomer. 
Conversely, unlike DEO-EA, the PMC polymer is more prone to decomposition when near the cis conformer.
This highlights the importance of considering initial configurations in predicting decomposition pathways.
The findings contribute to a deeper understanding of the stability and kinetics of polymer electrolyte materials, offering valuable 
insights for the design of advanced solid-state batteries.

\section*{Acknowledgements}
The authors thank the German Federal Ministry of Education and Research (BMBF) for financial support through the FestBatt2 project 
(Grant Number 03XP0435E). Computations simulations were performed on the PALMA-II HPC cluster of the University of M\"unster.

%\section*{References}

\bibliography{references_abr}

\end{document}

%\setchemformula{bond-length=1em}

\begin{frontmatter}

	\title{Insights into Polymer Electrolyte Stability and Reaction Pathways: A first-principle calculations study\tnoteref{mytitlenote}}

	\author[affil:KZ]{Kazem Zhour\corref{mycorrespondingauthor}}
	\ead{kazem.zhour@uni-muenster.de}
	\address[affil:KZ]{Institut für Physikalische Chemie, Universit\"at M\"unster, Correnssta{\ss}e 28, 48149 M\"unster, Germany}
	\cortext[mycorrespondingauthor]{Corresponding authors}
	
	\author[affil:DD]{Diddo Diddens}
	\address[affil:DD]{Helmholtz Institute M\"unster (IEK-12), Forschungszentrum J\"ulich GmbH, Corrensstra{\ss}e 46, 148159 M\"unster}
	\ead{d.diddens@fz-juelich.de}
	
	\author[affil:KZ]{Andreas Heuer}
	\ead{andheuer@uni-muenster.de}

\end{frontmatter}

\section{Supporting information}

\begin{figure}[!ht]
	\centering
	\includegraphics[width=0.5\linewidth]{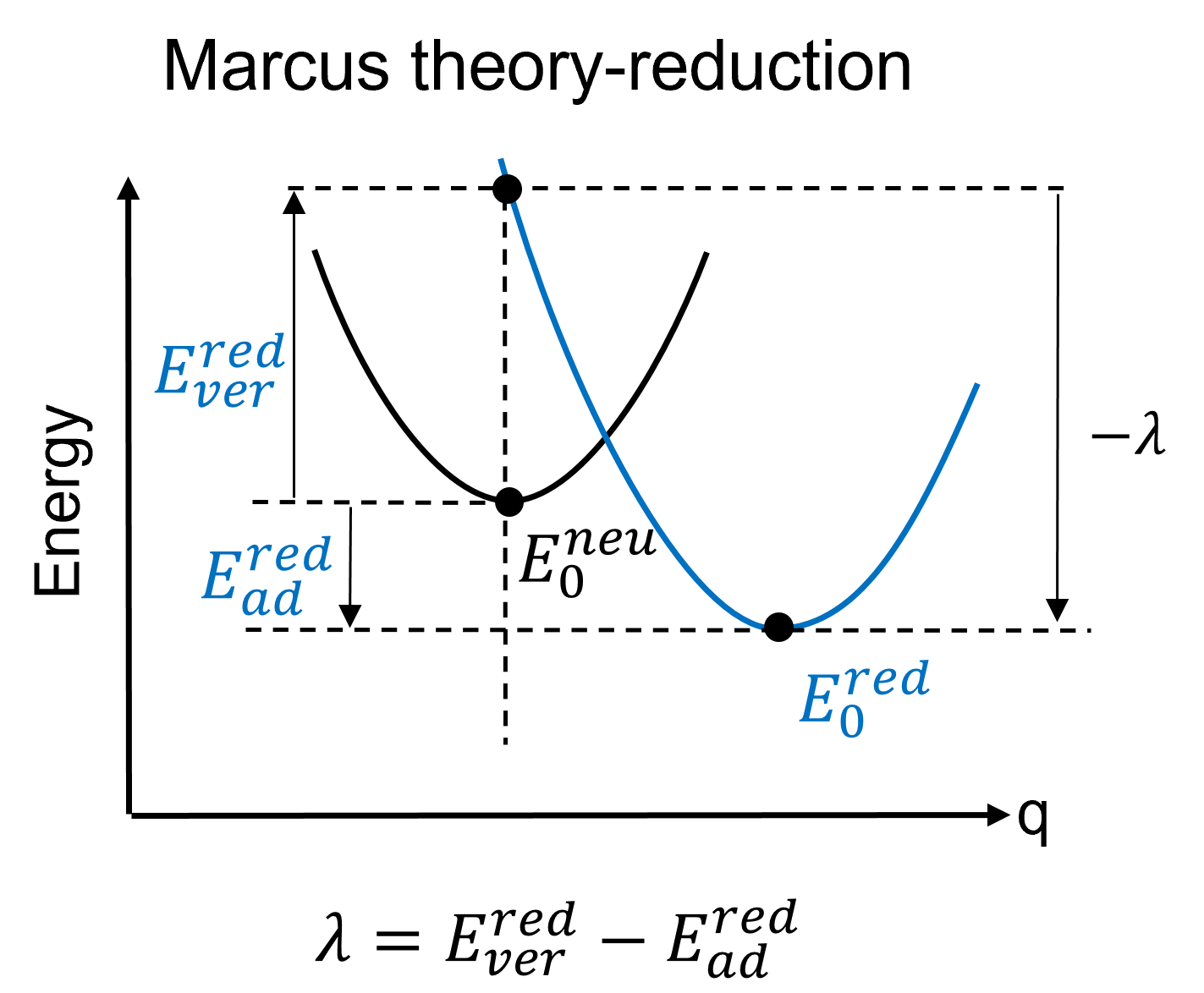}
	\caption{Reorganization energy ($\lambda$) calculated from Marcus theory of oxidation, featuring the 
	vertical reduction energy ($E^{red}_{ver}$) and the adiabatic reduction energy ($E^{ox}_{ad}$).}
	\label{figS:marcus-red}
\end{figure}

\begin{figure}[!ht]
	\centering
	\includegraphics[width=0.5\linewidth]{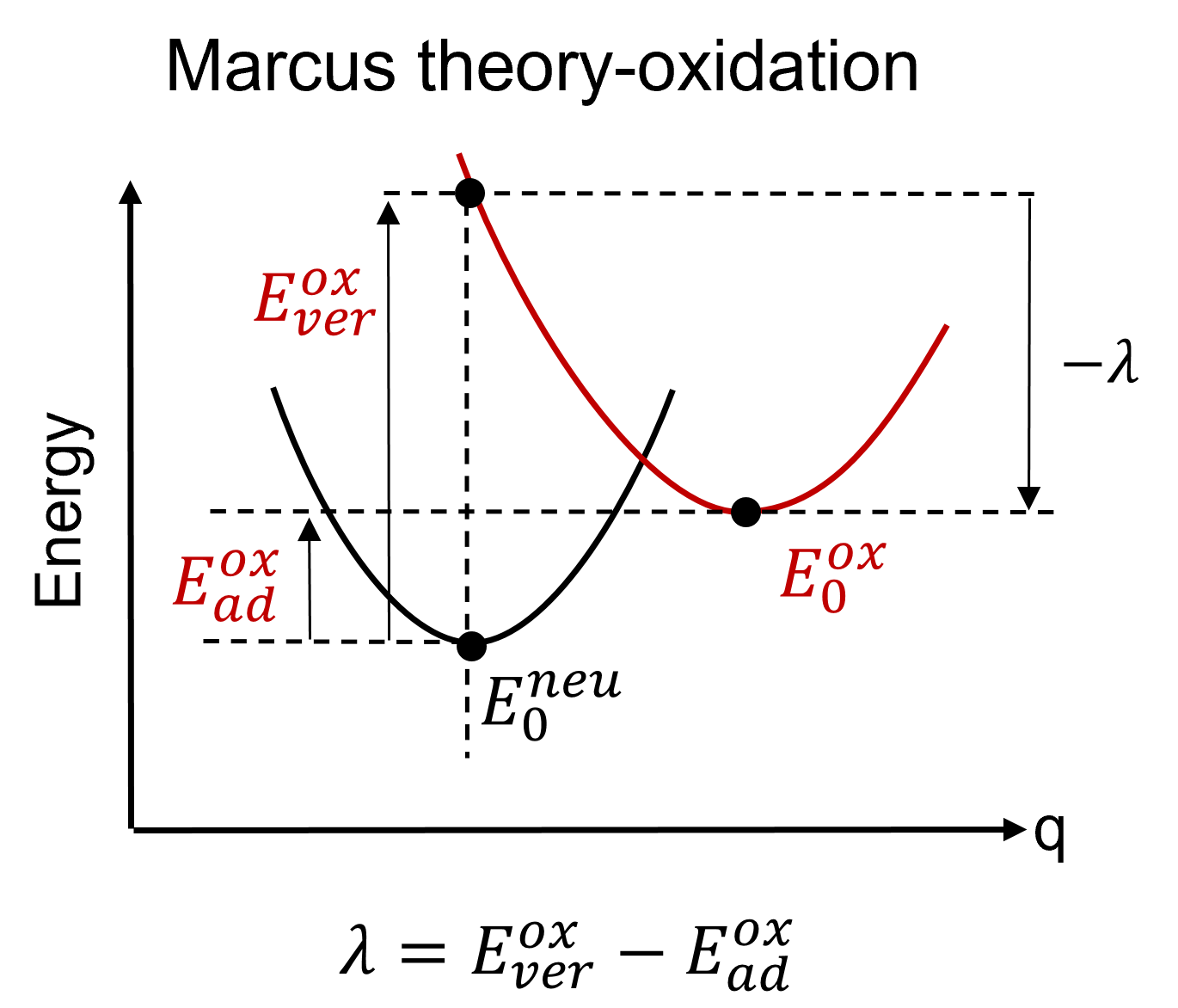}
	\caption{Reorganization energy ($\lambda$) calculated from Marcus theory of oxidation, featuring the 
	vertical oxidation energy ($E^{ox}_{ver}$) and the adiabatic oxidation energy ($E^{ox}_{ad}$).}
	\label{figS:marcus-oxi}
\end{figure}

\begin{table}[h!]
\caption{\label{tab:dihedral_TFSI}
Dihedral angle of [TFSI]$^-$ in degree for neutral, oxidized, and reduced monomers with and without [Li]$^+$.}
\smallskip
\begin{center}
\begin{tabular}{lc*{1}{r@{.}l}c*{1}{r@{.}l}c*{1}{r@{.}l}}
\hline
         && \multicolumn{8}{c}{d} \\
		 \cline{3-10}
cluster  && \multicolumn{2}{c}{neutral} && \multicolumn{2}{c}{reduced} && \multicolumn{2}{c}{oxidized} \\
\hline
PMC[TFSI]$^-$             &&  171&18  &&  178&70  &&  158&23  \\
PMC[LiTFSI]               &&  168&45  &&  167&92  &&  167&51  \\
PMC-OH[TFSI]$^-$          &&  177&25  &&  178&81  &&  160&48  \\
PMC-OH[LiTFSI]            &&  163&17  &&  161&59  &&  159&42  \\
PeMC-OH[TFSI]$^-$         &&  175&43  &&  173&58  &&  174&78  \\
PeMC-OH[LiTFSI]           &&  172&38  &&  173&57  &&  169&04  \\
DEO-EA[TFSI]$^-$          &&   30&85  &&   85&95  &&   17&65  \\
DEP-EA[LiTFSI]            &&   37&47  &&   42&62  &&   38&23  \\
\hline
\end{tabular}
\end{center}
\end{table}

\begin{figure}[!ht]
	\centering
	\includegraphics[width=\linewidth]{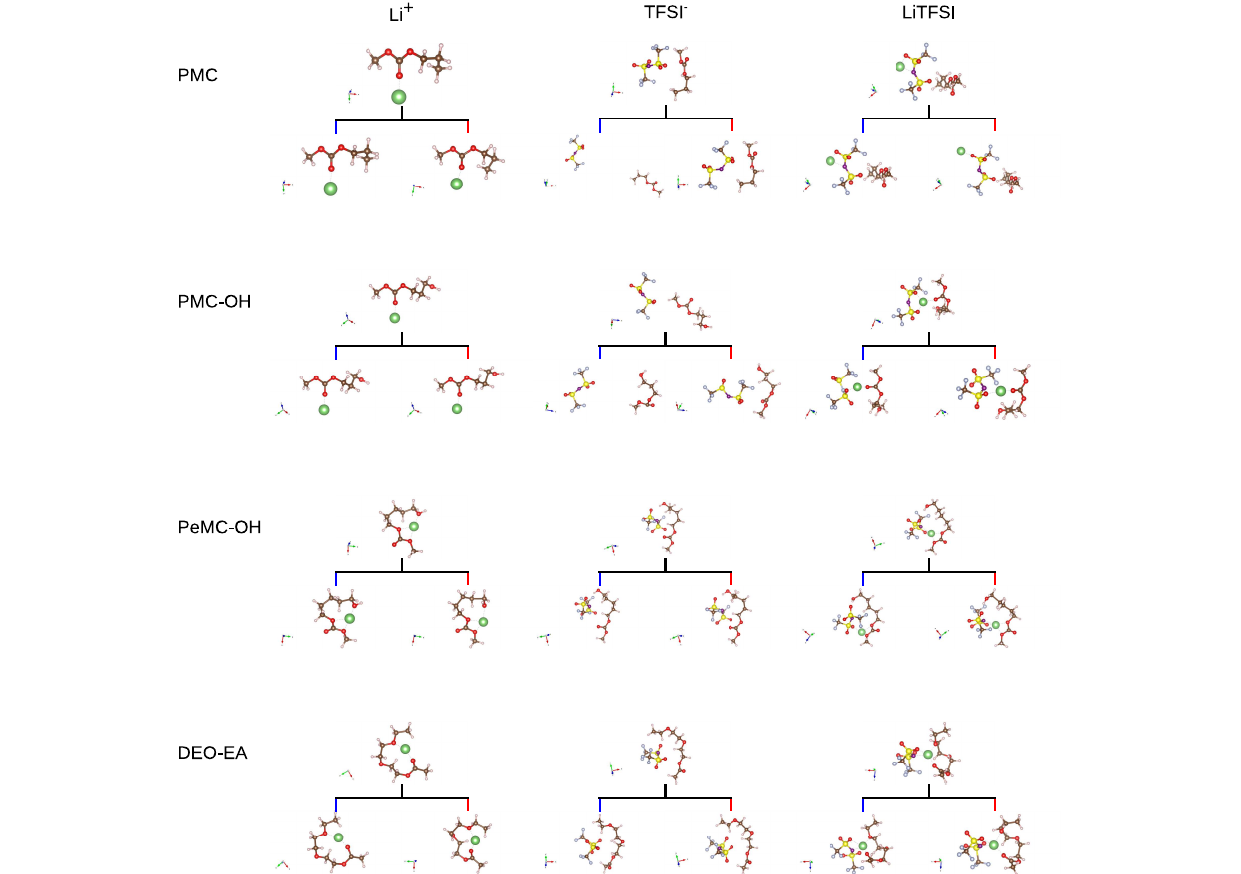}
	\caption{Structural representations of neutral (top, centered), reduced (below blue lines), and oxidized (below red lines) 
	states for PMC, PMC-OH, PeMC-OH, and DEO-AE monomers in proximity to [Li]$^+$, [TFSI]$^-$, and [LiTFSI].
	The color code of the atoms is brown for carbon, red for oxygen, white for hydrogen, green for lithium, 
	purple for nitrogen, yellow for sulfur, and grey for florine.}
	\label{figS:oxi-red}
\end{figure}

\begin{figure}[!ht]
	\centering
	\includegraphics[width=0.5\linewidth]{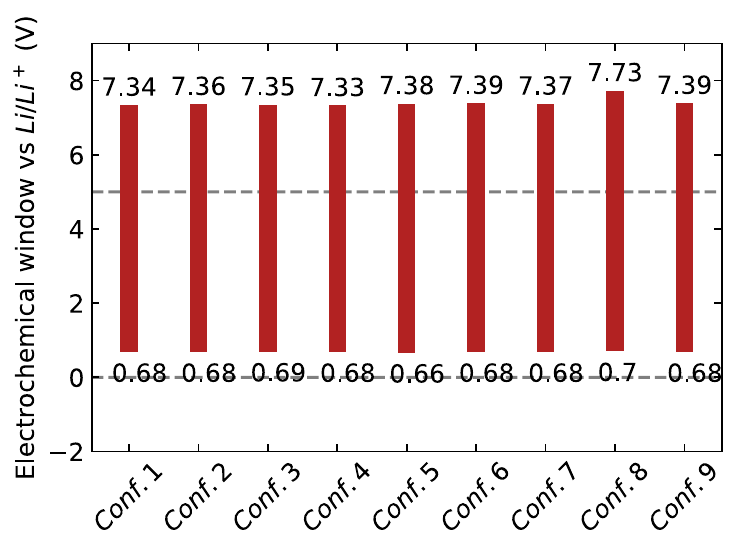}
	\caption{\label{figS:confs}
	EWs of the first nine most stable conformers of PMC near [Li]$^+$. 
	The dashed lines at 0 and 5 V correspond to the battery's EW.
	The upper limits of the histogram represent the oxidation potential ($V_{\rm red}$) 
	while the lower ones denote the reduction potential ($V_{\rm ox}$).}
\end{figure}

\begin{figure}[!ht]
	\centering
	\includegraphics[width=0.75\linewidth]{figures/confs.pdf}
	\caption{The first nine most stable conformers of PMC near Li cation, generated using CREST software.
	The color code is similar to Fig. \ref{figS:oxi-red}.}
	\label{figS:confs_configs}
\end{figure}

\begin{figure}[!ht]
	\centering
	\includegraphics[width=\linewidth]{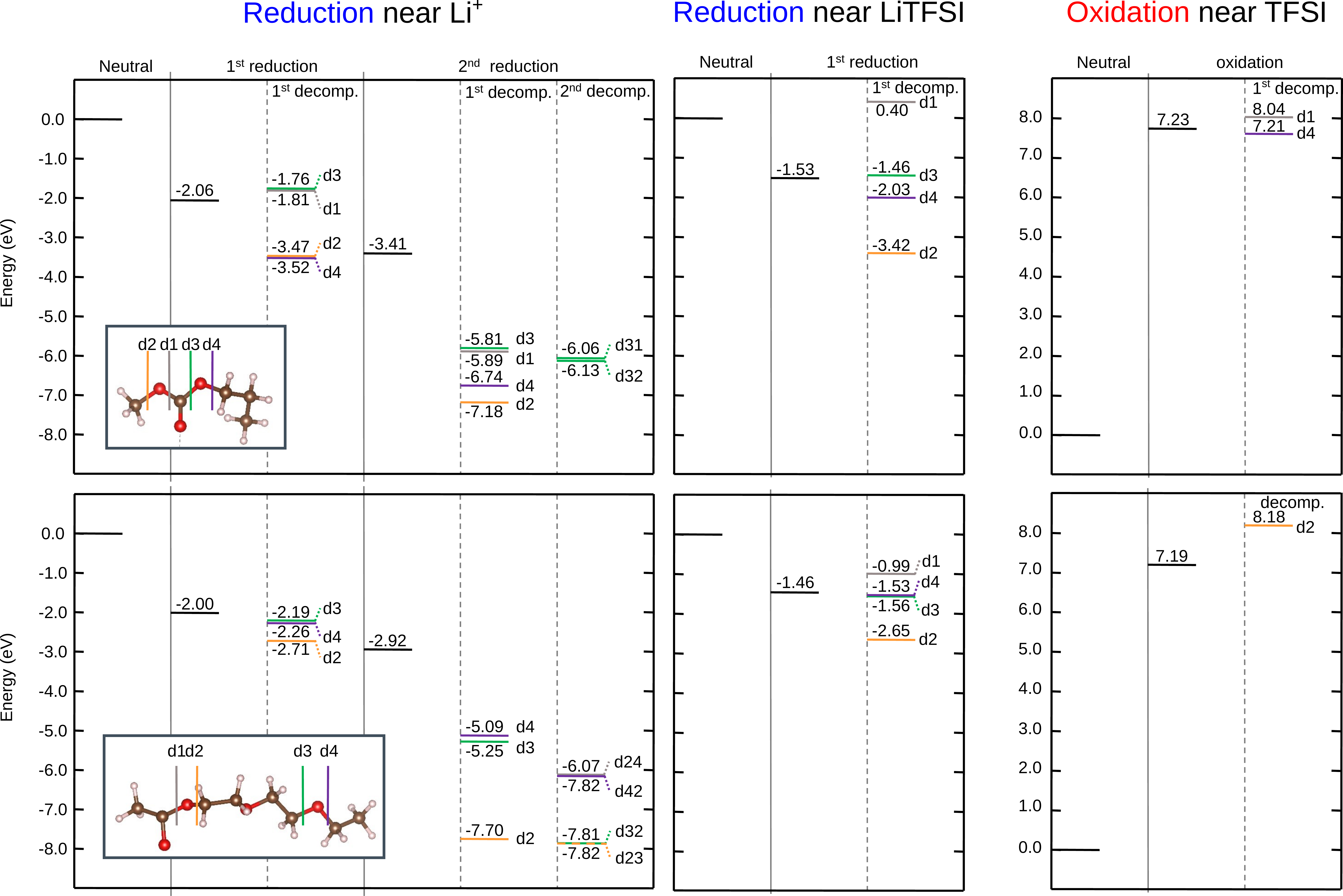}
	\caption{\label{fig:decomp}
	Energy diagram of the decomposition of PMC (top diagrams) and DEO-EA (bellow diagrams) after first and second reduction reduction near [Li]$^+$, 
	first reduction near [Li$]^+$[TFSI]$^-$, and oxidation near [TFSI]$^-$.
	The color code is similar to Fig. \ref{figS:oxi-red}.}
\end{figure}

\begin{figure}[!ht]
	\centering
	\includegraphics[width=\linewidth]{figures/F1_Li.pdf}
	\caption{Decomposition pathway of PMC near Li ion.
	The color code is similar to Fig. \ref{figS:oxi-red}.}
	\label{figS:F1_Li}
\end{figure}

\begin{table}[h!]
\caption{\label{tab:F1_Li}
Distance (d in \AA) for each coordinated atom from the Li ion.}
\smallskip
\begin{center}
\begin{tabular}{lc*{1}{r@{.}l}c*{1}{r@{.}l}c*{1}{r@{.}l}}
\hline
         && \multicolumn{8}{c}{d} \\
		 \cline{3-10}
cluster  && \multicolumn{2}{c}{1} && \multicolumn{2}{c}{2} && \multicolumn{2}{c}{3} \\
\hline
PMC[Li]$^+$                 &&  1&76  &&  &    &&     &     \\
PMC$^-$[Li]$^+$             &&  1&85  &&  &    &&     &     \\
PMC$^-$d$_1$[Li]$^+$        &&  1&66  && 1&94  &&     &     \\
PMC$^-$d$_1^-$[Li]$^+$      &&  1&71  && 1&95  &&     &     \\
PMC$^-$d$_2$[Li]$^+$        &&  1&93  && 1&92  &&     &     \\
PMC$^-$d$_2^-$[Li]$^+$      &&  2&04  && 2&01  &&    2&10  \\
PMC$^-$d$_3$[Li]$^+$        &&  1&96  && 1&69  &&     &     \\
PMC$^-$d$_3^-$[Li]$^+$      &&  1&82  && 1&71  &&     &     \\
PMC$^-$d$_3^-$d$_1$[Li]$^+$ &&  1&73  && 1&73  &&     &     \\
PMC$^-$d$_3^-$d$_2$[Li]$^+$ &&  2&11  && 1&72  &&     &     \\
PMC$^-$d$_4$[Li]$^+$        &&  1&94  && 1&93  &&     &     \\
PMC$^-$d$_4^-$[Li]$^+$      &&  2&05  && 2&01  &&    2&11  \\
\hline
\end{tabular}
\end{center}
\end{table}

\begin{figure}[!ht]
	\centering
	\includegraphics[width=\linewidth]{figures/F1_LiTFSI.pdf}
	\caption{Decompostion pathway of PMC near LiTFSI ion.}
	\label{figS:F1_LiTFSI}
\end{figure}

\begin{table}[h!]
\caption{\label{tab:F1_LiTFSI}
Number (N) of coordinated atoms to the Li ion, and corresponding distance (d in \AA) for each coordinated atom. Dihedral angle of TFSI in degree.}
\smallskip
\begin{center}
\begin{tabular}{lc*{1}{r@{.}l}c*{1}{r@{.}l}c*{1}{r@{.}l}c*{1}{r@{.}l}}
\hline
         && \multicolumn{8}{c}{d} & \multicolumn{3}{c}{$\phi$} \\
		 \cline{3-10}
cluster  && \multicolumn{2}{c}{1} && \multicolumn{2}{c}{2} && \multicolumn{2}{c}{3} & \\
\hline
PMC[LiTFSI]                   &&  1&99  &&  2&06  &&      &    &&  168&45        \\
PMC$^-$[LiTFSI]               &&  2&10  &&  2&27  &&      &    &&  167&91        \\
PMC$^-$d$_1$[LiTFSI]          &&  1&98  &&  2&06  &&      &    &&  166&53        \\
PMC$^-$d$_2$[LiTFSI]          &&  1&89  &&  1&90  &&     1&98  &&  168&10        \\
PMC$^-$d$_3$[LiTFSI]          &&  2&12  &&  2&22  &&     1&74  &&  159&97        \\
PMC$^-$d$_4$[LiTFSI]          &&  1&99  &&  2&06  &&      &    &&  171&14        \\
\hline
\end{tabular}
\end{center}
\end{table}

\begin{figure}[!ht]
	\centering
	\includegraphics[width=\linewidth]{figures/F1_TFSI.pdf}
	\caption{Decompostion pathway of PMC near TFSI ion.
	The color code is similar to Fig. \ref{figS:oxi-red}.}
	\label{figS:F1_TFSI}
\end{figure}

\begin{table}[h!]
\caption{\label{tab:F1_TFSI}
Number (N) of coordinated atoms to the Li ion, and corresponding distance (d in \AA) for each coordinated atom. Dihedral angle of TFSI in degree.}
\smallskip
\begin{center}
\begin{tabular}{lc*{1}{r@{.}l}}
\hline
cluster &  \multicolumn{3}{c}{$\phi$} \\
\hline
PMC[TFSI]$^-$               &&  171&18        \\
PMC$^+$[TFSI]$^-$           &&  158&23        \\
PMC$^+$d$_1$[TFSI]$^-$      &&  166&96        \\
PMC$^+$d$_2$[TFSI]$^-$      &&      &          \\
PMC$^+$d$_3$[TFSI]$^-$      &&      &        \\
PMC$^+$d$_4$[TFSI]$^-$      &&   162&38        \\
\hline
\end{tabular}
\end{center}
\end{table}

\begin{figure}[!ht]
	\centering
	\includegraphics[width=\linewidth]{figures/F3_Li.pdf}
	\caption{Decomposition pathway of DEO-EA near Li ion.
	The color code is similar to Fig. \ref{figS:oxi-red}.}
	\label{figS:F3_Li}
\end{figure}

\begin{table}[h!]
\caption{\label{tab:F3_Li}
Distance (d in \AA) for each coordinated atom from the Li ion.}
\smallskip
\begin{center}
\begin{tabular}{lc*{1}{r@{.}l}c*{1}{r@{.}l}c*{1}{r@{.}l}c*{1}{r@{.}l}}
\hline
         && \multicolumn{10}{c}{d} \\
		 \cline{3-13}
cluster  && \multicolumn{2}{c}{1} && \multicolumn{2}{c}{2} && \multicolumn{2}{c}{3}  && \multicolumn{2}{c}{4} \\
\hline
DEO-EA[Li]$^+$                 &&  1&82  && 1&90  &&     &    &&     &    \\
DEO-EA$^-$[Li]$^+$             &&  1&71  && 1&92  &&     &    &&     &    \\
DEO-EA$^-$d$_2$[Li]$^+$        &&  1&91  && 1&91  &&     &    &&     &    \\
DEO-EA$^-$d$_2^-$[Li]$^+$      &&  2&05  && 2&06  &&    2&09  &&    1&85  \\
DEO-EA$^-$d$_2^-$d$_3$[Li]$^+$ &&  2&09  && 2&02  &&    1&85  &&     &    \\
DEO-EA$^-$d$_2^-$d$_4$[Li]$^+$ &&  2&12  && 2&04  &&    2&21  &&    2&12  \\
DEO-EA$^-$d$_3$[Li]$^+$        &&  1&88  && 1&68  &&     &    &&     &    \\
DEO-EA$^-$d$_3^-$[Li]$^+$      &&  2&05  && 1&80  &&    2&16  &&     &    \\
DEO-EA$^-$d$_3^-$d$_2$[Li]$^+$ &&  2&06  && 2&04  &&    1&74  &&     &    \\
DEO-EA$^-$d$_4$[Li]$^+$        &&  1&91  && 1&71  &&     &    &&     &    \\
DEO-EA$^-$d$_4^-$[Li]$^+$      &&  2&11  && 1&81  &&    2&17  &&     &    \\
DEO-EA$^-$d$_4^-$d$_2$[Li]$^+$ &&  1&93  && 1&98  &&    2&13  &&     &    \\
\hline
\end{tabular}
\end{center}
\end{table}

\begin{figure}[!ht]
	\centering
	\includegraphics[width=\linewidth]{figures/F3_LiTFSI.pdf}
	\caption{Decomposition pathway of DEO-EA near LiTFSI.
	The color code is similar to Fig. \ref{figS:oxi-red}.}
	\label{figS:F3_LiTFSI}
\end{figure}

\begin{table}[h!]
\caption{\label{tab:F3_LiTFSI}
Number (N) of coordinated atoms to the Li ion, and corresponding distance (d in \AA) for each coordinated atom. Dihedral angle of TFSI in degree.}
\smallskip
\begin{center}
\begin{tabular}{lc*{1}{r@{.}l}c*{1}{r@{.}l}c*{1}{r@{.}l}c*{1}{r@{.}l}c*{1}{r@{.}l}c*{1}{r@{.}l}}
\hline
         && \multicolumn{14}{c}{d} & \multicolumn{3}{c}{$\phi$} \\
            \cline{3-16}
cluster  && \multicolumn{2}{c}{1} && \multicolumn{2}{c}{2} && \multicolumn{2}{c}{3} && \multicolumn{2}{c}{4} && \multicolumn{2}{c}{5} & \\

\hline
DEO-EA[LiTFSI]                 &&  2&09  &&  2&14  &&   1&87  &&    1&95  &&     &    && 37&46        \\
DEO-EA$^-$[LiTFSI]             &&  2&17  &&  2&26  &&   1&76  &&    2&01  &&     &    && 42&62        \\
DEO-EA$^-$d$_1$[LiTFSI]        &&  2&00  &&  1&98  &&   1&76  &&     &    &&     &    && 44&12        \\
DEO-EA$^-$d$_2$[LiTFSI]        &&  2&20  &&  2&39  &&   2&02  &&    2&00  &&    1&99  && 34&20        \\
DEO-EA$^-$d$_3$[LiTFSI]        &&  1&99  &&  1&95  &&   1&73  &&     &    &&     &    && 39&98        \\
DEO-EA$^-$d$_4$[LiTFSI]        &&  2&10  &&  2&53  &&   1&97  &&    1&73  &&     &    && 24&40        \\
\hline
\end{tabular}
\end{center}
\end{table}

\begin{figure}[!ht]
	\centering
	\includegraphics[width=\linewidth]{figures/F3_TFSI.pdf}
	\caption{Decomposition pathway of DEO-EA near TFSI ion.
	The color code is similar to Fig. \ref{figS:oxi-red}.}
	\label{figS:F3_TFSI}
\end{figure}

\begin{table}[h!]
\caption{\label{tab:F1_TFSI}
Dihedral angle of TFSI in degree.}
\smallskip
\begin{center}
\begin{tabular}{lc*{1}{r@{.}l}}
\hline
cluster &  \multicolumn{3}{c}{$\phi$} \\
\hline
DEO-EA[TFSI]$^-$               &&  30&85        \\
DEO-EA$^+$[TFSI]$^-$           &&  66&74        \\
DEO-EA$^+$d$_1$[TFSI]$^-$      &&    &          \\
DEO-EA$^+$d$_2$[TFSI]$^-$      &&  38&81        \\
DEO-EA$^+$d$_3$[TFSI]$^-$      &&     &         \\
DEO-EA$^+$d$_4$[TFSI]$^-$      &&     &         \\
\hline
\end{tabular}
\end{center}
\end{table}

\begin{figure}[!ht]
	\centering
	\includegraphics[width=0.5\linewidth]{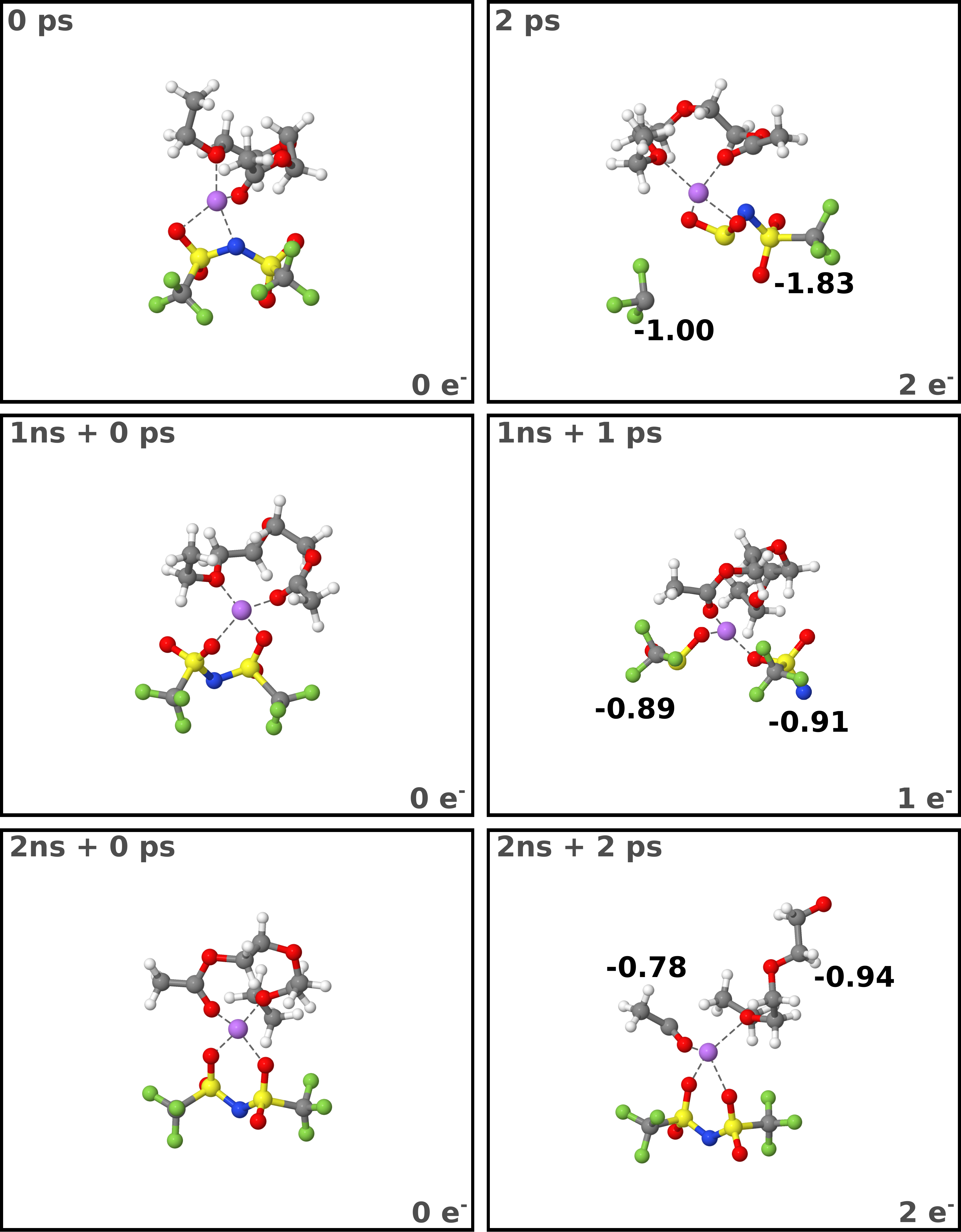}
	\caption{Diagrams of first decomposition of DEO-AE near [Li$]^+$[TFSI]$^-$. The initial configuration at the top is provided by the best 
	conformers according to CREST calculations, while the two subsequent configurations are obtained after 1 ns of consecutive molecular 
	dynamics based on the extended tight-binding force field. AIMD simulations run for 1 ps after adding each electron.
	Charges of the produced fragments are given in |e|.
	The dashed line shows the coordination of the atoms in the system with Li.
	The color code is: grey for carbon, red for oxygen, white for hydrogen, purple for lithium, blue for nitrogen, yellow for sulfur, 
	and green for fluorine.}
	\label{figS:AIMD_F3_LiTFSI}
\end{figure}